# Tensile strained In$_x$Ga$_{1-x}$P membranes for cavity optomechanics

Garrett D. Cole,[1,a)] Pen-Li Yu,[2] Claus Gärtner,[1] Karoline Siquans,[1] Ramon Moghadas Nia,[1,3] Jonas Schmöle,[1] Jason Hoelscher-Obermaier,[1,3] Thomas P. Purdy,[2] Witlef Wieczorek,[1] Cindy A. Regal,[2] Markus Aspelmeyer[1]

[1]*Vienna Center for Quantum Science and Technology (VCQ), Faculty of Physics, University of Vienna, A-1090 Vienna, Austria*
[2] *JILA, National Institute of Standards and Technology and University of Colorado, Boulder, Colorado 80309, USA and Department of Physics, University of Colorado, Boulder, Colorado 80309, USA*
[3] *Max-Planck-Institute for Gravitational Physics, Albert-Einstein-Institute, and Institut für Gravitationsphysik, Leibniz Universität Hannover, Callinstraße 38, D-30167 Hannover, Germany*

We investigate the optomechanical properties of tensile-strained ternary In$_x$Ga$_{1-x}$P nanomembranes grown on GaAs. This material system combines the benefits of highly strained membranes based on stoichiometric silicon nitride, with the unique properties of thin-film semiconductor single crystals, as previously demonstrated with suspended GaAs. Here we employ lattice mismatch in epitaxial growth to impart an intrinsic tensile strain to a monocrystalline thin film (approximately 30 nm thick). These structures exhibit mechanical quality factors of $2\times10^6$ or beyond at room temperature and 17 K for eigenfrequencies up to 1 MHz, yielding Q*f products of $2\times10^{12}$ Hz for a tensile stress of ~170 MPa. Incorporating such membranes in a high finesse Fabry-Perot cavity, we extract an upper limit to the total optical loss (including both absorption and scatter) of 40 ppm at 1064 nm and room temperature. Further reductions of the In content of this alloy will enable tensile stress levels of 1 GPa, with the potential for a significant increase in the Q*f product, assuming no deterioration in the mechanical loss at this composition and strain level. This materials system is a promising candidate for the integration of strained semiconductor membrane structures with low-loss semiconductor mirrors and for realizing stacks of membranes for enhanced optomechanical coupling.

Cavity optomechanics is a rapidly evolving field operating at the intersection of solid-state physics, modern optics, and materials science [1]. The fundamental process at the heart of this interdisciplinary endeavor is the enhancement of radiation pressure within a high-finesse optical cavity. Isolating this weak interaction, specifically the momentum transfer of reflecting photons, requires the development of high-performance mechanical resonators that simultaneously exhibit excellent optical quality (requiring low absorption and scatter loss) and minimal mechanical dissipation. In recent years, a diverse suite of solutions have been developed, covering nearly 20 orders of magnitude in effective mass and 10 orders of magnitude in frequency [2]. In many cases, advances in semiconductor-derived micro- and nanofabrication processes have been the driving force behind the improvement in device performance. An effective implementation of such systems, which enables separate optimization of the optical cavity and mechanical resonator, employs nanomembranes dispersively coupled to the optical mode of a high-finesse Fabry-Perot cavity. Such membranes consist of extremely thin (<100 nm) dielectric [3] or semiconducting films [4], with lateral dimensions in the millimeter range. Dielectric structures typically comprise tensile-strained amorphous films of stoichiometric silicon nitride deposited by LPCVD [5,6]. Such membranes exhibit a mechanical quality factor (Q) in excess of $10^7$ [3,5,7,8] and have enabled a number of exciting developments in the

---
[a)] Electronic mail: garrett.cole@univie.ac.at

field of cavity optomechanics, including the demonstration of radiation-pressure quantum backaction [9] and ponderomotive squeezing [10]. In a similar vein, semiconductor nanomembranes based on free-standing epitaxial GaAs have shown excellent optomechanical quality [4] leading to the demonstration of exciton-mediated photothermal cooling [11,12]. The unique electro-optic properties of compound semiconductors enable coupling to quantum-electronic systems such as quantum wells [13] and quantum dots [14]. However, given the narrow bandgap of GaAs (1.42 eV), this material is limited to transparent operation for wavelengths longer than about 870 nm at room temperature and is subject to nonlinear absorption effects, including two-photon absorption (TPA), at high optical intensities.

In this letter we investigate an alternative material choice that combines the positive attributes of $Si_3N_4$, i.e. the ability to tune the resonator stress state, with the unique electro-optic properties of compound semiconductors, with the potential for improved optical transparency in the near infrared. We explore a tensile-strained single-crystal nanomembrane realized via lattice mismatch in epitaxial growth, through variations of the alloy composition of a ternary $In_xGa_{1-x}P$ (InGaP) layer. Fabricating suspended membranes from a nominally 30-nm thick film, we record room temperature and cryogenic mechanical quality factors at or beyond $2\times10^6$ for eigenfrequencies up to 1 MHz. Furthermore, through cavity-mediated measurements of the membrane optical properties, we find a combined scatter and absorption loss of 40 ppm for these membranes at 1064 nm and 300 K.

Epitaxial InGaP thin films are commonly employed in microwave transistors owing to the superior electronic properties (band alignment with GaAs, high electron saturation velocity, etc.) [15], as well as their unique chemical properties, namely the potential for selective etching with respect to GaAs and AlGaAs compounds [16]. The latter property has also led to the use of InGaP as a sacrificial material in the development of micromechanical resonators [17]. High quality thin films can be realized on GaAs substrates via epitaxial growth processes such as molecular beam epitaxy (MBE) or metal organic chemical vapor deposition (MOCVD). Beyond microelectronics, the direct bandgap of 1.9 eV (~650 nm) of this compound leads to wide use in photonic applications, including the production of red-emitting laser diodes (both surface and edge emitters) [18], LEDs [19], and multi-junction solar cells [20]. The stress state of these ternary alloy films can be tailored through variations in the group-III composition leading to deviations from the ideal lattice matching conditions with the underlying GaAs substrate. Thus, an InGaP surface layer can be grown tensile strained (In content <49%), strain free (lattice matched, $In_{0.49}Ga_{0.51}P$), or compressive (In content >49%). While extensively employed in micro- and optoelectronic applications, InGaP has not previously been used as a mechanical material. Thus, the mechanical dissipation of InGaP thin films remains unexplored.

Previous efforts have examined the optomechanical characteristics of various unstrained compound-semiconductor-based resonators [4,21-27], while select works have focused on the benefits of tensile strain in these systems [28,29]. Tension has proven to be an effective means for boosting the Q*f product, particularly in nitride-based resonators [30-33]. Beneficially, intrinsic stress increases the resonant frequencies of the membrane, while keeping the mechanical linewidth approximately constant [33]. We extend these previous developments to lattice-mismatched InGaP on GaAs, a common and commercially-relevant epitaxial materials system that allows for highly selective etching for ease of microfabrication. A further advantage of InGaP, when compared with previously demonstrated GaAs nanomembranes, is the potential for high transparency in the near infrared. Due to its relatively wide bandgap, the TPA coefficient of lattice-matched InGaP is only 7.8 cm/GW at 1064 nm [34], much lower than that of GaAs at approximately 20 cm/GW [35]. Moreover, at 1550 nm, TPA is completely suppressed in InGaP owing to its 1.9 eV bandgap [36]. At high optical intensities this significantly reduces undesired photothermal effects and heating of the membrane, which is crucial when operating such devices at ultralow temperatures. Finally, the high refractive index of InGaP compounds (3.22 at 1064 nm and room temperature compared to approximately 2 for $Si_3N_4$ under the same conditions) leads to an increase in optomechanical coupling for a dispersively-coupled intra-cavity membrane.

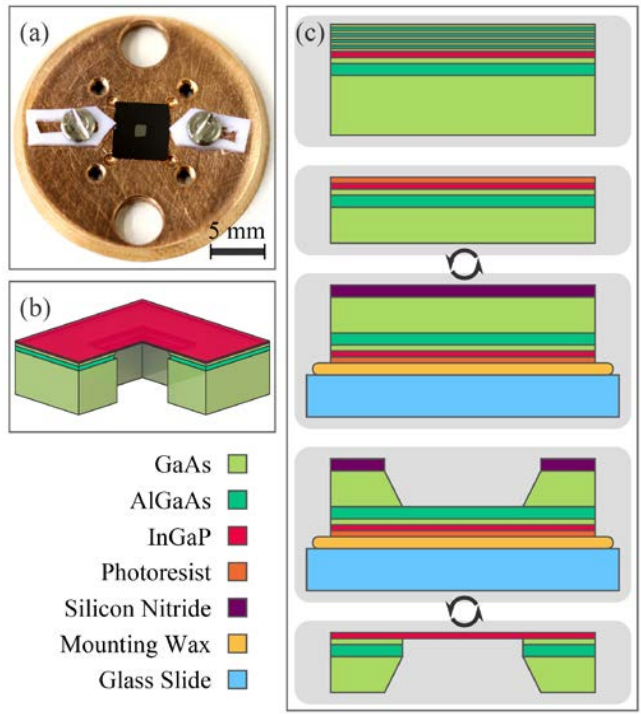

Fig.1: Tensile-strained InGaP membrane development. (a) Photograph of a completed InGaP membrane (nominal size of 1×1 $mm^2$) chip clamped lightly to a copper mount. (b) Cross-sectional solid model of a free-standing 30-nm thick InGaP membrane structure indicating the final layer structure. (c) Schematic of select steps for the membrane microfabrication procedure; see details in text.

As shown in Figure 1, our monocrystalline membranes are fabricated from an epitaxial multilayer consisting originally of a high-reflectivity multilayer mirror atop a nominally 30-nm thick InGaP film, which in turn lies above a double etch stop structure comprised of GaAs and high-aluminum-content AlGaAs. This structure is grown by MBE on a 150-mm diameter, 675-um thick, semi-insulating (100) GaAs substrate. Note that both the Bragg mirror and dual-etch-stop layers are not required for this work; in this case the design was chosen as it existed "off the shelf." Future structures will simply consist of a surface InGaP layer and will eliminate (or at least minimize the thickness of) the underlying GaAs film. Previous work has demonstrated a selectivity approaching $10^7$:1 for AlGaAs etching over InGaP in a dilute HF solution [37].

Fabrication of our devices (Fig. 1c) entails a single-mask bulk micromachining process beginning with the removal of the Bragg stack from the surface of the structure using a phosphoric-based etch chemistry ($H_3PO_4$:$H_2O_2$:$H_2O$, 1:5:15 volume ratio). After exposing the underlying InGaP layer, the growth substrate is thinned to 150-200 µm. During this process, the front-face of the wafer is protected with a cured photoresist layer. Following re-polish of the substrate, a backside silicon nitride hard mask is deposited via plasma-enhanced chemical vapor deposition. The thinned and nitride-coated samples are then attached to a temporary mount, consisting of a glass slide coated with high temperature wax. This temporary mount ensures structural rigidity of the thinned GaAs substrate and eventually the free-standing membrane. Lithographic patterning using standard optical contact lithography defines square windows of 0.5×0.5 mm$^2$ or 1×1 mm$^2$ on the backside of the substrate. These features will ultimately control the lateral extent of the resonator geometry.

The backside window pattern is transferred into the silicon nitride mask via a reactive ion etching process using $SF_6$. Following the hard mask etch step, the GaAs substrate is removed using a selective $H_2O_2$:$NH_4OH$ (30:1 volume ratio) wet chemical etch. In this case we place the sample in an ultrasonic bath while etching to enhance the uniformity of the substrate removal process. This process terminates on the high-Al content etch stop layer, which is stripped with dilute HF. This step exposes the underlying GaAs film, which is in turn etched away with the same phosphoric acid chemistry used to remove the Bragg mirror. This process shows sufficient selectivity over InGaP in order to generate the desired free-standing 30-nm thick membranes. As mentioned above, this GaAs layer is unnecessary for the development of the devices. However, there may be some advantages in terms of material quality in growing InGaP on a GaAs surface rather than directly atop a high-Al content AlGaAs film. To complete the processing of the resonators, the samples are soaked in acetone to separate them from the temporary mount and the membrane chips are gently dried after a thorough solvent cleaning.

It is important to note that in contrast to KOH-etching of silicon, as is commonly employed to produce $Si_3N_4$ membranes, the semi-isotropic nature of the GaAs substrate removal process makes it difficult to accurately control the lateral geometry of the resonators; in many cases the membrane shape deviates from the ideal square geometry of the backside mask pattern (Fig. 1a). Fortunately, as will be described below, these non-ideal membrane geometries still yield impressive mechanical quality factors. Moreover, even with the perturbed shape, the difference between the measured and calculated eigenfrequencies, generated from a simple formula for a rectangular membrane, are less than 2.5% for the nominally 1×1 mm$^2$ InGaP membranes (Fig. 2). Regardless, further work is required in order to achieve repeatability in the control of the device geometry.

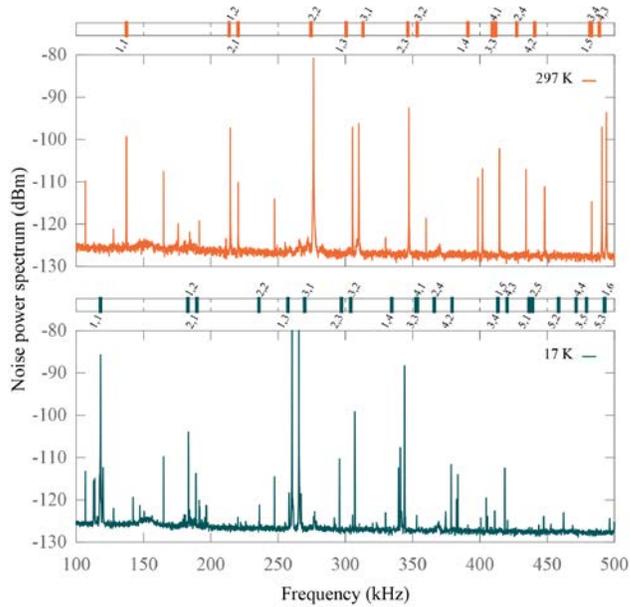

Fig. 2: Frequency response of a nominally 1×1 mm$^2$ InGaP membrane recorded with a low-noise optical homodyne interferometer at room (top) and low temperature (bottom). Here we plot the noise power spectrum out to approximately 500 kHz when exciting the resonator with a piezoelectric actuator driven with a white noise voltage signal. Modeling the geometry as a rectangle, we find a reasonable match between the measured and theoretical eigenmodes (see bar above each data set) for lateral dimensions of 0.92×0.97 mm$^2$.

Measurements of the membrane mechanical response are carried out in a custom fiber-optic interferometer with a $^4$He cryostat as a sample chamber, allowing characterization from room temperature to about 15 K [24]. Additional independent measurements are carried out in a room-temperature vacuum chamber in Boulder, Colorado, USA. In each case, the membranes are operated at a vacuum level below 10$^{-5}$ mbar in order to minimize viscous damping. We first record the optically probed mechanical displacement, as shown in Figure 2, to identify the membrane modes. To this end, the resonators are mechanically driven by applying a white-noise voltage signal to a piezoelectric transducer (piezo) fixed to the sample holder. The noise power spectrum of the mechanical motion reveals several clear mechanical eigenmodes. Fitting a selection of peaks and approximating the membrane shape as a rectangle, we extract lateral dimensions of 0.92×0.97 mm$^2$ for this nominally 1×1 mm$^2$ structure. From fitting of the frequency response of a number of membranes, both 0.5×0.5 mm$^2$ and 1×1 mm$^2$, we extract an in-plane tensile stress values of 150-170 MPa at room temperature for the InGaP layer. Note that in these samples the InGaP film was intended to be lattice matched; however, a slight deviation of -2% from the ideal In composition for the as-grown structure (corresponding to In$_{0.47}$Ga$_{0.53}$P) yields the observed stress level.

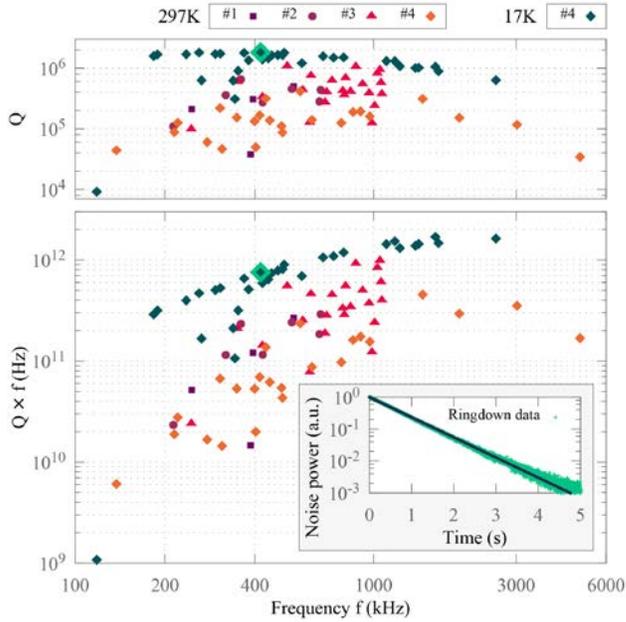

Fig. 3: Mechanical quality factor of tensile-strained InGaP membranes. Top: Compiled Q values (measured at both room and cryogenic temperature) as a function of frequency for a series of four devices. We observe relatively constant Q values for the cryogenic dataset, out to a frequency of ~1 MHz (equivalent mode number of 8,8). Bottom: Compiled Q*f product for the same devices. Inset: Example ringdown response for a nominally 1×1 mm$^2$ membrane at 17 K (4,3 eigenmode equivalent, 418 kHz). The extracted quality factor is $1.8 \times 10^6$.

In order to probe the mechanical dissipation of the structures (Fig. 3), select mechanical modes are individually driven by applying a short burst of a sinusoidally-varying voltage to the piezo at a chosen eigenfrequency, *f*. The amplitude decay time (1/e), $\tau$, of the mechanical ringdown yields the mechanical quality factor via $Q=\pi f \tau$. We record Q values as high as 2.7 million for a nominally 1×1 mm$^2$ membrane at room temperature (322 kHz resonance frequency, 3,2 eigenmode equivalent, $\tau$=2.64 s) and nearly 2 million at cryogenic temperature (17 K) for modes out to ~1 MHz (Fig. 3), yielding a maximum Q*f product of ~$2 \times 10^{12}$ Hz.

The optical loss of the membrane is estimated by recording the modulation of finesse due to the presence of the intra-cavity membrane [6,38]. Here, we insert the membrane in an open-air Fabry-Perot cavity with a bare finesse of $F_0$=30,600 at 1064 nm and an optical waist of 60 µm. For this measurement the membrane is translated along the cavity axis using a piezoelectric actuator with a step size of 50 nm. After each step, one mirror of the cavity is scanned quickly and repeatedly to simulate periodic delta-function excitations of the cavity field. The exponential intensity decay is recorded and used to extract the photon lifetime within the cavity. The finesse varies from $0.35F_0$ to $1.2F_0$ as the membrane is translated over one optical wavelength. These extrema can be extracted despite a cavity performance that is observed to be somewhat unstable at the positions of maximum dispersive coupling. The minimum finesse value is compared with a theoretical model to obtain an upper bound for the total optical loss. Measurements obtained from two distinct positions on a 1×1 mm$^2$ membrane yield an estimate of the total optical loss, including both absorption and scatter, of 40 ppm.

Optical inspection of the membranes using white-light microscopy reveals the presence of macroscopic defects in or on the suspended structure. These defects will lead to excess scatter and thus limit our current loss estimates to being conservative upper limits for this material. Most likely these scatterers arise as a consequence of the additional etching step required to strip the nearly 7-µm thick Bragg mirror from the surface of the membrane. Future samples, consisting solely of the InGaP membrane layer, will likely yield lower scatter losses through the elimination of these defects. Post-processing analysis of the membrane surface quality by atomic force microscopy yields an RMS microroughness value of 0.3 nm for $10\times10$ µm$^2$ scan area, corresponding to a scatter loss of 12 ppm at 1064 nm. Subtracting this from the total loss yields a value of 28 ppm that can be attributed to a combination of scattering from the aforementioned macroscopic defects or background optical absorption in the InGaP film. Attributing this remaining loss to absorption in the 30-nm membrane thickness, 28 ppm is equivalent to an absorption coefficient of ~10 cm$^{-1}$ at 1064 nm, corresponding to a conservative estimate of the imaginary component of the refractive index of $8\times10^{-5}$.

Looking ahead, we envision constructing optimized membrane structures with increased tension. This will be realized through further reductions in the In mole fraction to levels significantly below 49%. According to the critical thickness criterion developed in Ref. [39], the maximum strain that can be accommodated in a 300-Å thick InGaP layer on a GaAs substrate is 0.85%, corresponding to an In content of 36%. At this composition, the tensile stress of a rigidly clamped and non-relaxed membrane structure would be 1.1 GPa, using a biaxial modulus of 129 GPa for the In$_{0.36}$Ga$_{0.64}$P thin film. As the resonator frequencies scale with the square-root of the intrinsic tensile stress, this increased stress level corresponds to a nearly three-fold increase in the membrane eigenfrequencies, assuming similar geometries as studied here. Assuming no degradation in the mechanical quality factor, the Q*f product should at least increase in a corresponding manner. It is also important to note that these higher tension samples will have an even wider bandgap, exceeding 2 eV [19], potentially leading to further improvements in the optical transparency.

Further envisioned benefits of this materials system, beyond the respectable optomechanical performance outlined above, are the ability to realize direct integration of high-tension crystalline membranes with low-loss Bragg mirrors [40] as well as the possibility for the direct growth of vertically-stacked membranes. Such structures have been theoretically investigated for their potential to generate very strong optomechanical coupling [41] and with the incorporation of high reflectivity surface-normal photonic crystal reflectors [42], may ultimately have the potential for realizing single-photon strong coupling. Such epitaxially-grown membrane stacks would be inherently aligned and parallel, with excellent (tenth of a percent level) thickness control enabling precise separation distances. Finally, as described in the introduction, this unique direct-bandgap compound semiconductor system is amenable to integration with "active" quantum electronic elements [15,18-20].

We have demonstrated tensile-strained monocrystalline membranes based on epitaxial InGaP. These structures show Q values approaching 3 million at room temperature with

Q*f products of nearly $2\times10^{12}$ Hz at cryogenic temperatures for a tensile stress of just under 200 MPa. Changes in the alloy composition will enable an increase in stress to levels up to 1 GPa. Assuming similar dissipation rates can be maintained at this level of strain, we anticipate an enhancement in the Q*f product towards the $10^{13}$ Hz regime. The wide bandgap of this compound can in principle yield a small optical absorption; our current measurements indicate a total extracted optical loss of 40 ppm (including both scatter and absorption) at 1064 nm, translating to an imaginary component of the refractive index below $1\times10^{-4}$ with microroughness induced scattering removed. Future work will explore the integration of these structures with high-performance crystalline Bragg mirrors as well as the development of stacks of optomechanically-coupled membranes.

**Acknowledgments**
G.D.C. acknowledges Robert Yanka and colleagues from IQE North Carolina for growth of the epitaxial structure. The Boulder group thanks R. W. Peterson for insightful discussions. P.-L.Y. thanks the Taiwan Ministry of Education for support, while C.R. thanks the Clare Boothe Luce Foundation. W.W. acknowledges support from the European Commission through a Marie Curie Fellowship. Additional funding is provided by the EC via projects ITN cQOM and iQUOEMS, the Austrian Science Fund (FWF project I909, FWF Doctoral Program CoQuS W 1210), the European Research Council (ERC StG QOM), the Vienna Science and Technology Fund (WWTF) under project ICT12-049, and the US National Science Foundation under Grant No. 1125844. Microfabrication was carried out at the Zentrum für Mikro- und Nanostrukturen (ZMNS) of the Technische Universität Wien.